\documentclass[twocolumn,showpacs,prb]{revtex4}
\usepackage{graphics,german,amsmath,amssymb}

\usepackage{graphicx}
\usepackage{color}
\usepackage{epstopdf}
\usepackage{bm}

\newcommand{\mvec}[1]{{\bf #1}}
\newcommand{\vl}[1]{\textcolor{black}{#1}}

\definecolor{olive}{rgb}{0.0,0.5,0.0}
\definecolor{darkred}{rgb}{0.5,0.0,0.0}

\begin{document}

\title{Tuning the magnetic ground state of a novel tetranuclear Nickel(II) molecular complex by high magnetic fields}
\author{C. Golze$^{1,2}$}
\author{A. Alfonsov$^{1,3}$}
\author{R. Klingeler$^{1,4}$}
\author{B. B"uchner$^{1}$}
\author{V. Kataev$^{1,3}$}
\author{C. Mennerich$^{2}$}
\author{H.-H. Klauss$^{2}$}
\author{M. Goiran$^{4}$}
\author{J.-M. Broto$^{4}$}
\author{H. Rakoto$^{4}$}
\author{S. Demeshko$^{5}$}
\author{G. Leibeling$^{5}$}
\author{F. Meyer$^{5}$}
\affiliation{$^{1}$Leibniz-Institute for Solid State and Materials Research IFW Dresden, P.O. Box 270116, D-01171 Dresden, Germany}

\affiliation{$^{2}$ Institut f\"ur Physik der Kondensierten Materie, TU Braunschweig, Mendelssohnstr.3, D-38106 Braunschweig, Germany}

\affiliation{$^{3}$ Kazan Physical Technical Institute of the Russian Academy of Sciences, 420029 Kazan, Russia}

\affiliation{$^{4}$Laboratoire National des Champs Magn\'{e}tiques Puls\'{e}s, 31432 Toulouse Cedex 04, France}

\affiliation{$^{5}$ Georg-August-University G"ottingen,  Institute of Inorganic Chemistry, D-37077 G"ottingen, Germany}
\date{\today}

\begin{abstract}
Electron spin resonance and magnetization data in magnetic fields up to 55\,T of a novel multicenter paramagnetic
molecular complex [L$_2$Ni$_4$(N$_3$)(O$_2$C\,Ada)$_4$](Cl\,O$_4)$ are reported. In this compound, four Ni centers each
having a spin  $S\,=\,1$ are coupled in a single molecule via bridging ligands (including a $\mu_4$-azide) which
provide paths for magnetic exchange. Analysis of the frequency and temperature dependence of the ESR signals yields the
relevant parameters of the spin Hamiltonian, in particular the single ion anisotropy gap and the $g$ factor, which
enables the calculation of the complex energy spectrum of the spin states in a magnetic field. The experimental results
give compelling evidence for tuning the ground state of the molecule by magnetic field from a nonmagnetic state at
small fields to a magnetic one in strong fields owing to the spin level crossing at a field of $\sim\,25$\,T.
\end{abstract}

\maketitle

\section{Introduction}

Novel quantum phenomena occurring in molecular magnets, such as quantum tunnelling of magnetization or spin-state transitions, as well as possible
future applications in quantum computing and data storage have attracted much interest to the interdisciplinary field of molecular magnetism (for
recent reviews see e.g. Ref.~\onlinecite{Barbara99,MM00,Verdaguer01,Gatteschi03}). Owing to the flexibility of synthetic chemistry a large variety of
complex molecular entities comprising a steadily increasing number of paramagnetic metal-ions have been synthesized. One important objective of this
synthesis is to increase the spin multiplicity of the molecular clusters as well as their magnetic anisotropy in order to increase their capabilities
for magnetic data storage. Regarding the goal of tailoring highly anisotropic molecular magnets, Nickel(II) is an attractive candidate as it often
exhibits a large single ion anisotropy. It is therefore increasingly used for the composition of molecular magnetic entities (see e.g.
Ref.~\onlinecite{Edwards03,delBarco04}). Recently a novel class of tetranuclear Nickel(II) complexes have been synthesized in which four Ni ions are
coupled in a single molecule via an unprecedented $\mu_4$-1,1,3,3 azide bridge.~\cite{Demeshko05} The azide ion is a preferred bridging ligand since
it can mediate different types of strong magnetic coupling depending on its coordination mode. \cite{Ribas99} Therefore, the structural flexibility
of the quadruply bridging azides brings about diverse molecular structures with different geometries of Ni$_4$ arrays and yields significant
variations in magnetic properties.

Information on magnetically driven spin excitations in magnetic molecular complexes is provided by magnetization studies in high magnetic fields
which directly probe additional contributions to the susceptibility. In order to study the rich spectrum of spin states in these complexes, electron
spin resonance (ESR) is another potentially very useful local method because of the possibility to excite selectively spin-flip transitions between
different energy levels. However, the complexity of the energy spectrum which results from the high spin multiplicity and large magnetic anisotropy
often renders conventional ESR instrumentation incapable for such studies. In particular, paramagnets comprising Ni$^{2+}$ (3d$^8$, $S$=1) spin
centers are often silent to standard ESR techniques confined to a few frequencies below 100\,GHz and magnetic fields not exceeding several Tesla.
This is due to the absence of Kramers degeneracy of the Ni$^{2+}$ spin states which results in an appreciable splitting of the spin levels by the
effect of the crystalline electric field and the spin-orbit coupling in zero magnetic field (zero field splitting). \cite{AB} Therefore the microwave
energy quantum $h\nu$ at conventional ESR frequencies of $\nu\, = \, 10$\,GHz (X-Band), 36\,GHz (Q-Band) or 95\,GHz (W-Band) is too small to induce
the resonance transitions between the spin levels of Ni$^{2+}$ in the accessible magnetic field range. Recent technological advances in generation of
the microwave radiation up to 1\,THz and higher combined with the growing availability of strong magnetic fields, however, have opened a new research
frontier in the study of magnetic systems by means of tunable high-field ESR-spectroscopy (see, e.g. Ref.~\onlinecite{workshop02}). In particular, a
number of strongly correlated quantum magnets comprising Ni$^{2+}$ spins have been studied by this technique in recent years (see, e.g.
Ref.~\onlinecite{Edwards03,delBarco04,Brill94,Barra99,Pardi00,Sieling00,Ohta01,Manaka01,Hagiwara03}).

In this paper we report a detailed ESR study at frequencies up to 1\,THz and in magnetic fields up to 40\,T of a
representative of the novel class of Ni$_4$-molecular complexes with central $\mu_4$-azide bridge of the type
[L$_2$Ni$_4$(N$_3$)(O$_2$C\,Ada)$_4$](Cl\,O$_4)$ (NiAz).~\cite{Demeshko05} Complementary measurements of the static
magnetization have been performed in fields up to 55\,T. By mapping the frequency/magnetic field diagram of the
resonance modes and by measuring the temperature dependence of the ESR intensities we are able to estimate all relevant
parameters of the spin Hamiltonian which describes the low-energy spin states of the Ni$_4$-magnetic cluster. In
particular we find an appreciable {magnetic} anisotropy gap in the first excited $S\,=\,1$ multiplet of the complex
amounting to ${\Delta_1}\,=\,6.7$\,K. The calculated field dependences of the spin states suggest that strong magnetic
fields allow to tune the ground state from a non-magnetic $S\,=\,0$ to a magnetic $S\,=\,1$ state. This is confirmed by
our pulsed magnetic field ESR and magnetization data which provide direct evidence for a magnetic field induced
switching of the ground state to a strongly magnetic one at $H\,\approx\,25$\,T.
Using the experimentally determined $g$ factor and { the anisotropy gap} a reanalysis of the low field magnetic
susceptibility ~\cite{Demeshko05} gives more accurate values for the three isotropic magnetic exchange constants in
this complex.

\section{Experimental}

\begin{figure}
    \includegraphics[width=0.9\columnwidth]{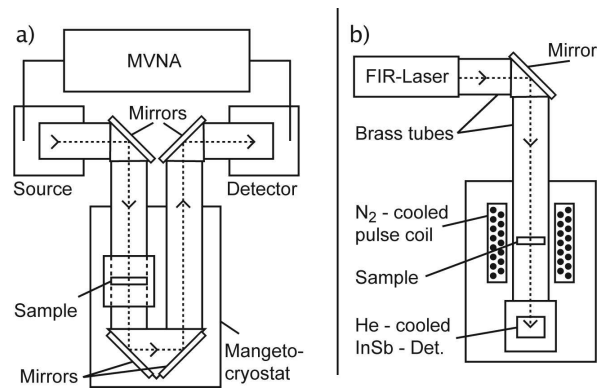}
    \caption{Experimental scheme for ESR measurements  in static (a) and pulsed magnetic fields (b), respectively. See text for details.}
    \label{Aufbau}
\end{figure}

For ESR measurements in static magnetic fields we used a Millimeterwave Vector Network Analyzer (MVNA) from AB
Millimetre, Paris, for generation of millimeter- and submillimeter microwaves and phase locked detection of a
signal.~\cite{Dahl98} The particular setup utilizes brass tubes as oversized waveguides and four 90$^{\circ}$ metallic
mirrors to lead the radiation into a superconducting cryomagnet to a sample positioned in the centrum of the
superconducting solenoid  and back to the detector (Fig.\ref{Aufbau}a). With this measurement scheme ESR in the
transmission mode has been measured in a frequency range from 20\,GHz up to 500\,GHz, in magnetic fields up to 15\,T
and at temperatures between 1.8 and 300\,K. ESR experiments in pulsed magnetic fields up to 40\,T have been performed
at the LNCMP Toulouse.~\cite{Portugall01} Microwaves in the frequency range of several hundred GHz to more than $1$ THz
{were obtained} from a Fabry-Perot cavity optically pumped by a CO$_2$ laser. Brass tubes and a 90$^{\circ}$ metallic
mirror are used to lead the radiation to the sample which sits in the center of the nitrogen cooled pulsed coil
(Fig.\ref{Aufbau}b). A field pulse with a duration of 800\,ms (raising time 120\,ms) is generated by a discharge of a
10\,kV capacitor bank on the coil. The radiation transmitted through the sample is detected by an InSb detector cooled
to liquid helium temperature. The sample temperature can be varied between 4.2 and 300\,K.

Complementary measurements of the static magnetization have been performed in pulsed magnetic fields up to 55\,T with a
pulse duration of 250\,ms (raising time 40\,ms) by an inductive technique. A system of two concentric pick-up coils in
opposition to each other equipped with additional compensation coils was utilized. The measured signal $\partial
M/\partial t$ was integrated numerically. The sample temperature can be varied between 1.6\,K and 300\,K.

\begin{figure}
    \includegraphics[width=0.9\columnwidth]{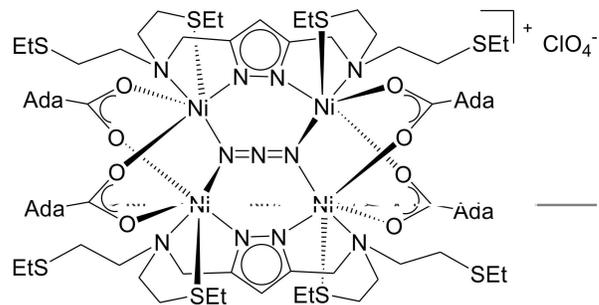}
    \caption{Molecular structure of Ni$_4$-complex [L$_2$Ni$_4$(N$_3$)(O$_2$C\,Ada)$_4$](Cl\,O$_4$).
    The four Ni(II)-ions are connected by the central azide, the pyrazolate and the carboxylate bridges.
    See Ref.~\onlinecite{Demeshko05} for details.}
    \label{Structure}
\end{figure}

Details of the synthesis and characterization of the compound under study (NiAz) can be found in
Ref.~\onlinecite{Demeshko05}. Its molecular structure is shown in Fig.~\ref{Structure}. For the ESR measurements, a
powder sample has been pressed into a pellet. A {diphenyl-picryl-hydrazyl} (DPPH) marker attached to the sample has
been used for field calibration.

\section{Results}

\subsection{ESR in static fields}
\label{staticESR}

\begin{figure}
    \includegraphics[angle=-90,width=\columnwidth]{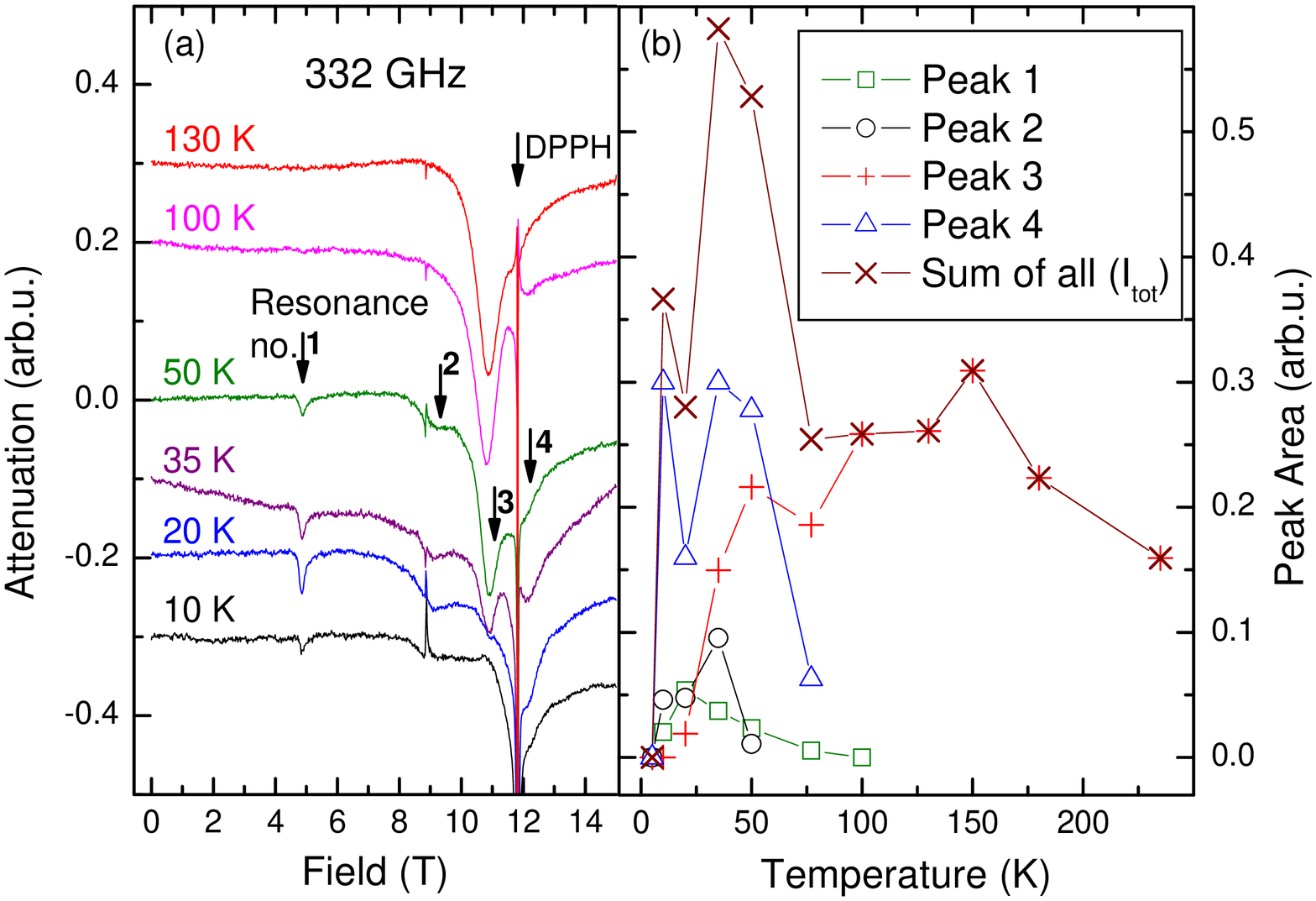}
    \caption{(Color online) (a) ESR spectra of NiAz at $\nu\,=\,332$\,GHz for different selected temperatures.
    Four thermally activated resonance peaks are indicated by the arrows and numbered from 1 to 4, respectively.
    A sharp resonance signal at 11.8\,T is the DPPH field marker.
    (b) Temperature dependence of the areas $I_i$ under the resonance peaks $i\,=\,1$ to 4 and the total area $I_{tot}$, respectively.}
    \label{TempDep}
\end{figure}

\begin{figure}
    \includegraphics[angle=-90,width=0.9\columnwidth,clip]{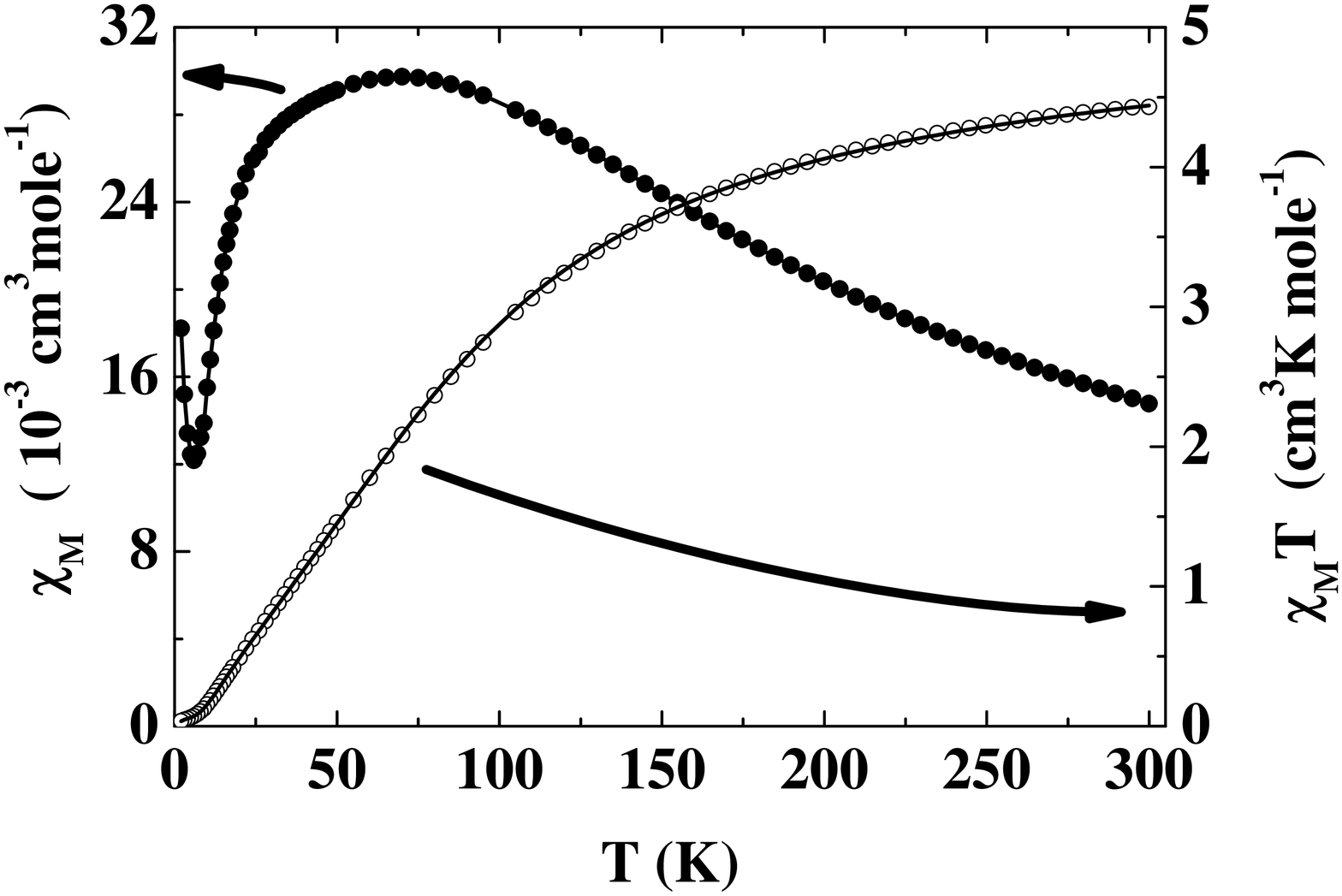}
    \caption{Magnetic susceptibility $\chi$ and $\chi T$ of NiAz measured in an external field of 0.5\,T,
    \vl{filled and open circles, respectively.}\cite{Demeshko05}
             The solid lines \vl{represent a numerical fit using the Hamiltonian (1). In this fit a temperature
             independent contribution $\chi_0\,=\,8.08\,\cdot\,10^{-4}$\,emu/mole as well as  a Curie-like impurity contribution corresponding
             to $\sim$\,2\,\% of ''defect'' Ni(II) paramagnetic sites which is responsible for a Curie tail below 20\,K are included.}}
\label{chi}
\end{figure}

At temperatures below 5\,K no ESR signals which can be attributed to the resonance of Ni$^{2+}$ have been detected in the static field measurements
at $H\,<\,15$\,T. However, a well-defined ESR spectrum develops with increasing temperature. In Fig.~\ref{TempDep}(a) representative spectra between
$10$\,K and $130$\,K at a fixed excitation frequency of $332$\,GHz are shown. Four resonances are observed by sweeping the magnetic field up to
$15$\,T, numbered from $1$ to $4$ in the order of their appearance in increasing field direction. The intensity of an ESR signal is proportional to
the static susceptibility of the resonating spins. \cite{AB} In Fig.~\ref{TempDep}(b) the temperature dependence of the individual resonance
intensities $I_i$, i.e. the areas under the respective absorption lines, and of the total intensity $I_{tot}(T)=\sum_i I_i$ are shown. Both $I_i(T)$
and $I_{tot}(T)$ exhibit an activated temperature behavior. In particular the total intensity $I_{tot}$ reveals a $T$ dependence similar to the
static susceptibility $\chi_{stat}$ results (see Ref.~\onlinecite{Demeshko05} and Fig.~\ref{chi}): when lowering the temperature from $300$ K it
shows a Curie-like paramagnetic increase, followed by a maximum around 50\,K. For lower temperatures, $I_{tot}$ goes to zero indicating a nonmagnetic
ground state. Examination of the temperature dependence of the intensities $I_i(T)$ of the individual resonances yields important additional
information. Peaks 1, 2 and 4 follow a common activation behavior with maxima around $30$ K, whereas resonance 3 shows a broad maximum at roughly
$150$ K [Fig.\ref{TempDep}(b)]. This observation suggests that resonances 1, 2 and 4 belong to resonance transitions in the same (first) excited spin
state, whereas, peak 3 is due to spin resonance in a higher lying energy state.

\begin{figure}
    \includegraphics[width=0.7\columnwidth]{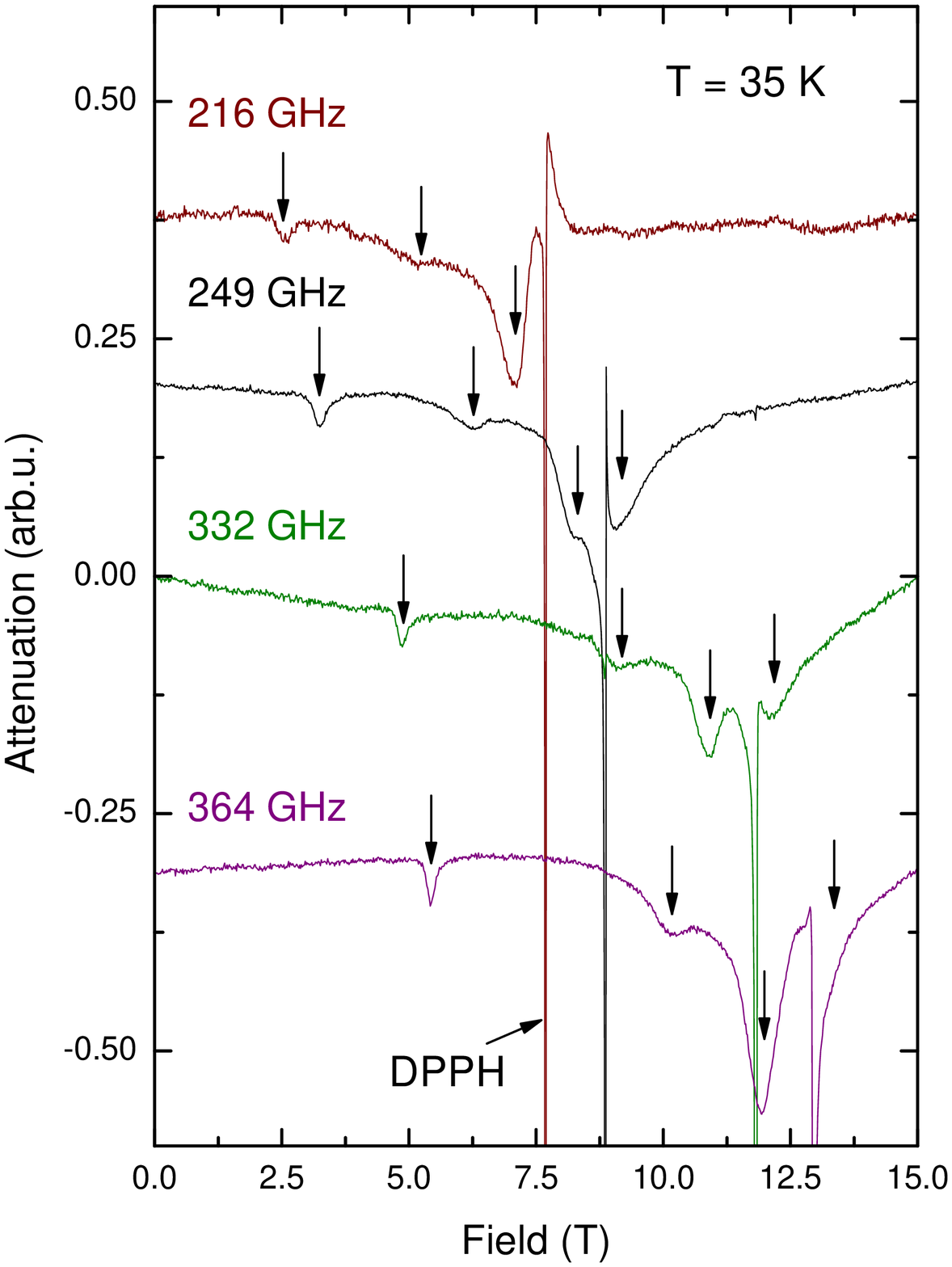}
    \caption{(Color online) ESR spectra of NiAz at $T\,=\,35$\,K at different selected excitation frequencies $\nu$.
    Four resonance modes are indicated by the arrows. Sharp lines are due to the DPPH field marker.}
\label{FreqDepSpec}
\end{figure}

\begin{figure}
    \includegraphics[width=0.7\columnwidth,angle=-90,clip]{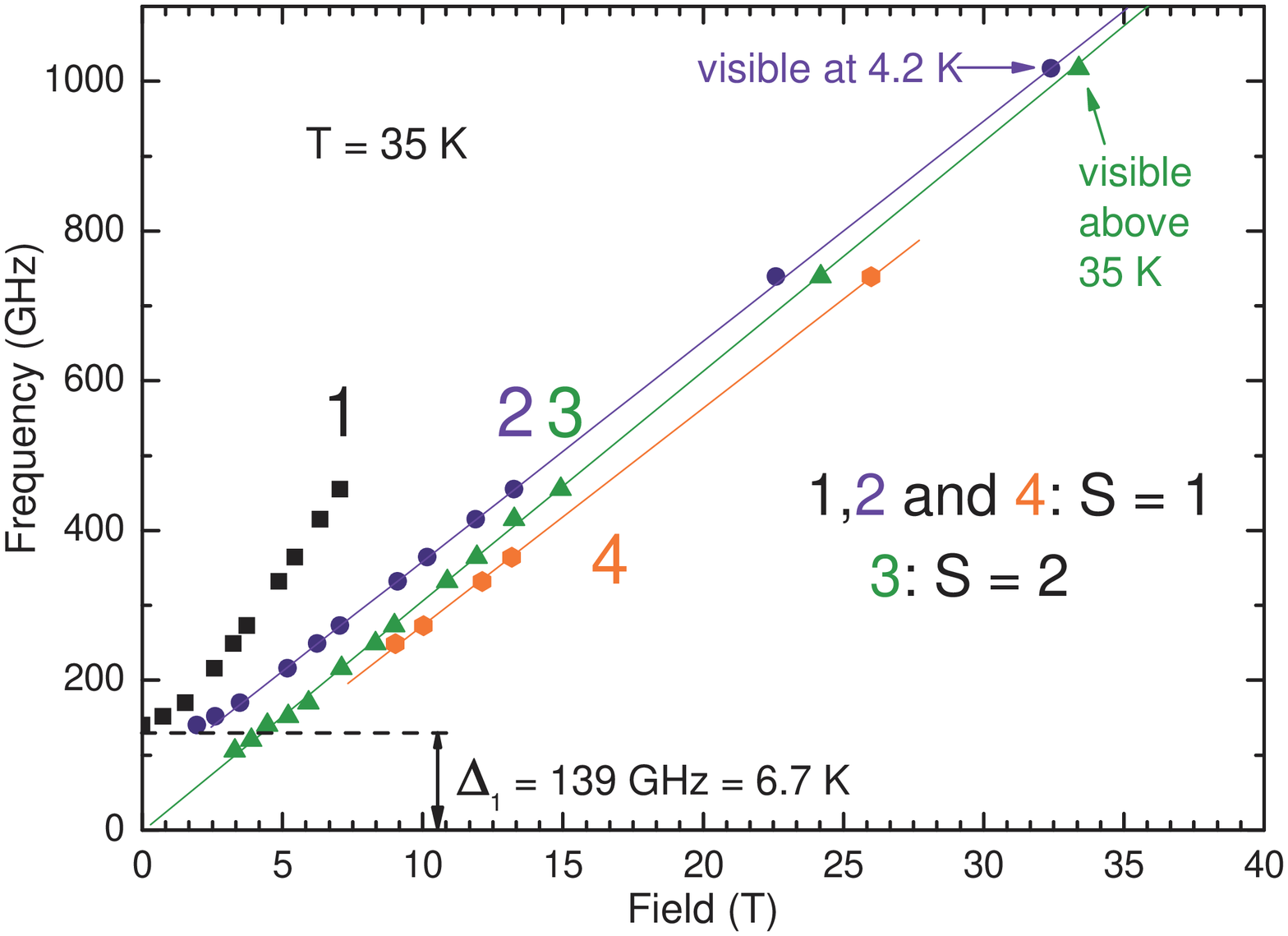}
    \caption{(Color online) Resonance branches at $T\,=\,35$\,K (including pulsed field ESR
    at $739$\,GHz ($35$\,K) and $1017$\,GHz ($4.2$\,K)). {Branches 1 and 2 have a finite frequency offset $\Delta_1$
    whereas branches 3 and 4 can be extrapolated to zero frequency at zero magnetic field}.
    Note that resonance 2 at 1017\,GHz is  visible at 4.2\,K owing to the spin level crossing.
    Solid lines are guide for the eye.}
\label{Branches}
\end{figure}

For a selected temperature of 35\,K at which all four resonance modes are clearly seen the ESR spectrum has been
measured at different excitation frequencies $\nu$. As can be seen in Fig.~\ref{FreqDepSpec} where several typical
spectra are plotted the position of the peaks in magnetic field $H^{res}_i$ depends on $\nu$, which ensures their
character as magnetic resonances. The relation between $\nu$ and $H^{res}_i$ is shown in Fig.~\ref{Branches}. The data
fall into four well defined resonance branches labelled from 1 to 4 according to the numbering of the detected
resonance modes. Remarkably, extrapolation of the branches 1 and 2 to zero magnetic field reveals that they intercept
the frequency axis at $\nu_0\,=\,139$\,GHz, i.e. there is a finite energy gap ${\Delta_1\,=\,}\nu_0h/k_B\,=\,6.7$\,K
for resonance excitations 1 and 2. On the other hand, branches 3 and 4 can be extrapolated to almost zero frequency at
zero field indicating that resonances 3 and 4 are gapless {within the experimental resolution}.

\subsection{ESR in pulsed magnetic fields}

\begin{figure}
    \includegraphics[width=0.9\columnwidth,clip]{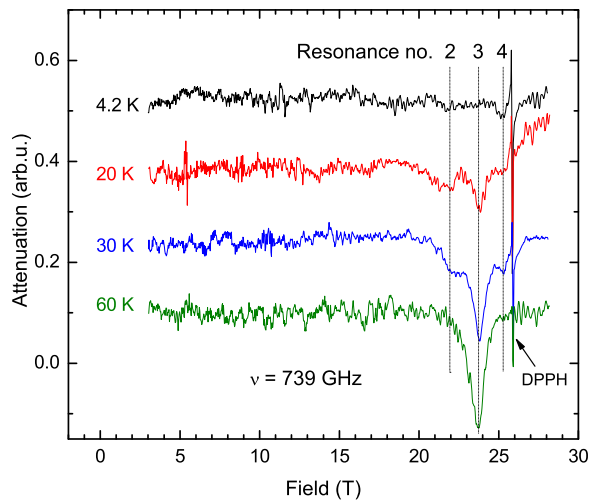}
    \caption{(Color online) ESR spectra of NiAz in pulsed magnetic field at 739\,GHz. Resonance modes 2, 3 and 4 indicated by thin lines
    are visible at elevated temperatures only.}
\label{PulseEPR075THzSpec}
\end{figure}

\begin{figure}
    \includegraphics[width=0.9\columnwidth,clip]{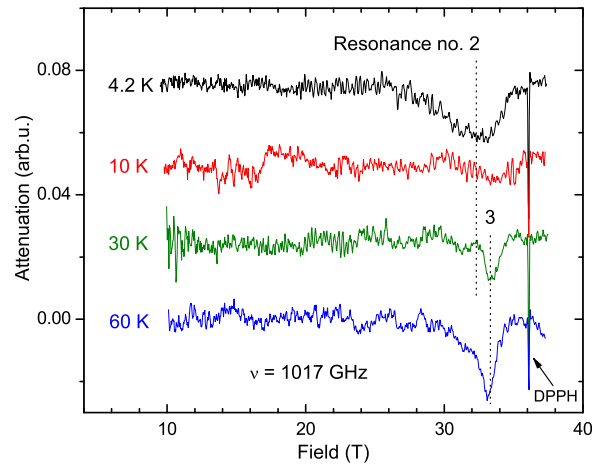}
    \caption{(Color online) ESR spectra of NiAz in pulsed magnetic field at 1017\, GHz.
    Note that at 4.2\,K a new broad line with a peak at about 32\,T appears in the spectrum.}
\label{PulseEPR1THzSpec}
\end{figure}

Measurements of ESR at an excitation frequency $\nu\,=\,739$\,GHz in pulsed magnetic fields confirm the static field data for smaller $\nu$.
Representative ESR spectra are shown in Fig.~\ref{PulseEPR075THzSpec}. No signals are detected at $T\,=\,4.2$\,K, whereas three thermally activated
resonances appear at $T\,\geq\,25$\,K. They can be straightforwardly assigned to the resonance modes 2, 3 and 4, respectively, as they nicely fall
into the respective $\nu$ vs. $H^{res}_i$ resonance branches (Fig.~\ref{Branches}). However, measurements of ESR at an excitation frequency
$\nu\,=\,1017$\,GHz yield a new feature in the spectrum: At a low temperature of 4.2\,K  a strong absorption mode develops at $H\,>\,25$\,T whose
intensity decreases with increasing $T$. At still higher temperatures another thermally activated resonance arises in the spectrum
(Fig.~\ref{PulseEPR1THzSpec}). The peak position $H^{res}_i$ of the former corresponds to branch 2, whereas the peak position of the latter - to
branch 3 (Fig.~\ref{Branches}).


\subsection{Magnetic Field Dependence of Magnetization}

\begin{figure}
    \includegraphics[width=0.7\columnwidth,clip,angle=-90]{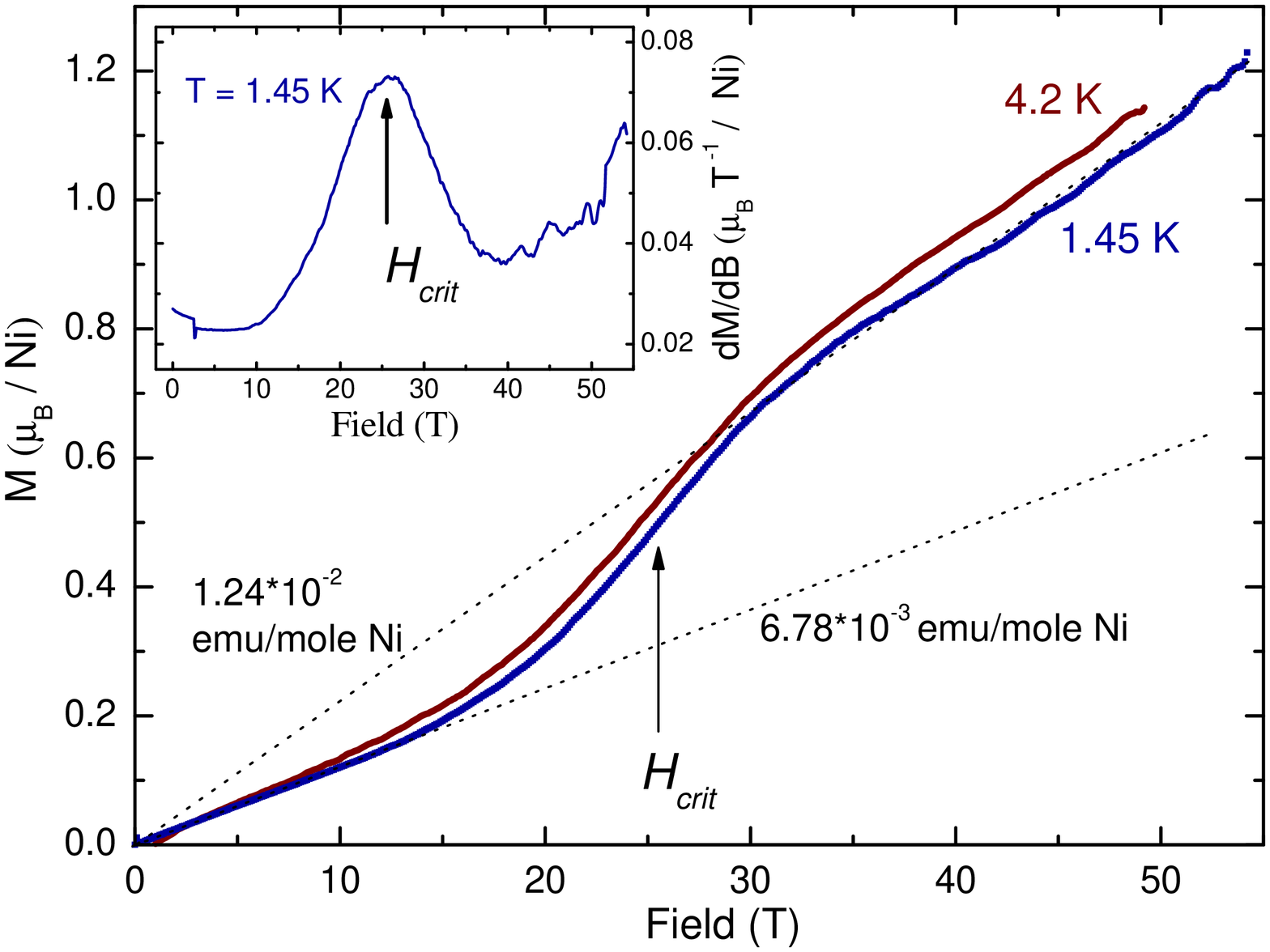}
    \caption{(Color online) Field dependence of the static magnetization $M(H)$ in pulsed magnetic field. Note a step like increase of $M$ at $\sim\,25$\,T and a
    linear background indicated by dot lines in the field range 0\,T\,-\,15\,T and 30\,T\,-\,50\,T, respectively.
    The field derivative of the magnetization $dM/dH$ for $T\,=\,1.45$\,K is shown in the inset.}
\label{PulsedFieldMagExp}
\end{figure}

The occurrence of the ESR resonance mode 2 already at a low temperature of 4.2\,K for $H\,>\,25$\,T should be related to changes in the static
magnetic properties of the sample. We hence performed magnetization studies $M(H)$ at 4.2\,K and 1.45\,K in pulsed fields up to 55\,T in order to
directly probe these changes. The $M(H)$ dependences at both temperatures are very similar and, indeed, the data yield a broad step-like increase of
$M(H)$ centered at a field $H_{crit}\,\sim\,25$\,T (Fig.~\ref{PulsedFieldMagExp}). \vl{The magnetization curves are reversible, i.e. no hysteresis
between the up- and down field sweeps can be observed within the experimental accuracy}. Remarkably, there is a finite slope of the magnetization at
$H\,<\,H_{crit}$ of the order $\partial M/\partial H\,=\,\chi\,=\,6.78\,\cdot\,10^{-3}$\, emu/mole~Ni which increases almost twice at
$H\,>\,H_{crit}$ reaching a value $\chi\,=\,1.24\,\cdot\,10^{-2}$\, emu/mole~Ni.

To summarize the experimental section, the central result of our measurements in pulsed magnetic fields concerns the fact that high frequency ESR
reveals an additional strong absorption mode at 4.2\,K in high magnetic fields above 25\,T. Concomitantly, the pulsed field magnetization exhibits a
broad step-like increase of $M(H)$, at low temperatures, centered at a field $H_{crit}\,\sim\,25$\,T. Both features signal drastic changes of the
magnetic ground state tuned by the magnetic field, which will be discussed in detail in the following section. Additionally, our experimental results
provide the diagram of the excitation frequency of the ESR modes versus resonance magnetic field. In the next section we will attribute these
features to specific spin transitions and, in particular, deduce important parameters like the $g$ factor and the zero field splitting (ZFS) gap. We
emphasize that the experimental determination of these quantities is essential for a comprehensive understanding of the spectrum of the spin states
of NiAz, specifically the occurrence of the magnetic field driven transition to a new magnetic state in strong fields.

\section{Discussion}

\subsection{The Effective Spin Hamiltonian: Minimal Model} \label{hamiltonian}

In order to address the magnetic properties of NiAz, in the following we discuss a minimal model to describe the magnetic field dependence of the
spin levels. In NiAz, the Ni$_4$-cores of each molecule are well separated in the solid phase. Hence, it is reasonable to assume that the
inter-molecular magnetic interactions are much weaker than the intra-molecular ones. The latter couplings are mediated via the azide bridge which
connects the four Ni centers with some contribution from the pyrazolate and carboxylate bridges (cf. Fig.~\ref{Structure}). The minimal effective
spin Hamiltonian to describe the magnetism of the Ni$_4$-complex should therefore contain the magnetic exchange interactions between the four
Ni-spins $S_{Ni}\,=\,1$. Additionally, the magnetic energy is determined by the single ion anisotropy caused by the crystal field (CF) and the
spin-orbit coupling as well as the Zeeman energy:

\begin{alignat*}{2}
    \ H \ = \   \phantom{+}&\,J_1 (\mvec{S}_1\cdot \mvec{S}_2 + \mvec{S}_3\cdot \mvec{S}_4)
            + \,J_2 (\mvec{S}_1\cdot \mvec{S}_3 + \mvec{S}_2\cdot \mvec{S}_4)\\
            +\,&J_3 (\mvec{S}_1\cdot \mvec{S}_4 + \mvec{S}_2\cdot \mvec{S}_3) \tag{1a}\label{exchange}\\
            +\,&D \sum_{i=1..4}\,[S_{iz}^2 - S_{Ni}(S_{Ni}+1)/3]\tag{1b}\label{anisotropy}\\
            +\,&g \mu_B \mvec{H}\cdot \sum_{i=1..4} \mvec{S}_i \tag{1c}\label{zeeman}
\end{alignat*}

\begin{figure}
    \includegraphics[width=0.5\columnwidth]{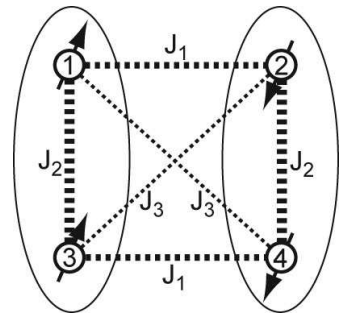}
    \caption{Proposed exchange paths between Ni spin centers in NiAz. See text for details.}
    \label{ExchangePaths}
\end{figure}

The first term (\ref{exchange}) of the Hamiltonian represents the isotropic exchange between the Ni-ions. The topology
of the Ni$_4$-core suggests three different exchange paths (see Fig.~\ref{ExchangePaths}), which are described by the
exchange parameters $J_{1,2,3}$. Here, $J_i\,>\,0$ denotes an antiferromagnetic (AF) coupling. The effect of the CF on
the Ni spins is accounted for by term (\ref{anisotropy}) in the spin Hamiltonian. In NiAz each Ni is surrounded by a
distorted octahedron comprising 3 nitrogen, 2 oxygen, and 1 sulfur ligands (Fig.~\ref{Structure}). Assuming that the
ligand CF is dominated by its axial component the interaction with the CF can be described by a single parameter $D$,
yielding term (\ref{anisotropy}) in the Hamiltonian. The magnitude of $D$ is determined by the average energy splitting
between the $3d$-orbital sets of the $t_{2g}$ and $e_g$ symmetry, respectively, $\Delta\,\sim\,1\,-\,2$\,eV, the
splitting within the $t_{2g}$ set, $\delta\,<<\,\Delta$, and the spin-orbit coupling $\lambda\,\simeq\,-40$\, meV,
$D\,\approx\,4\lambda^2\,\cdot\,(\delta/\Delta^2)$. \cite{AB} Note that in the case of $D\neq 0$ the energy of a
particular spin state depends on its magnetic quantum number $S_z$ even in zero magnetic field. This yields a ZFS gap
which {magnitude is proportional} to $D$. The third contribution to the Hamiltonian, (\ref{zeeman}), describes the
Zeeman interaction of the spins with the external magnetic field $\bf H$. In the case of uniaxial anisotropy, the $g$
factor reads $g_\parallel\,=\,2\,-\,8\lambda/\Delta$ and $g_\perp\,=\,2\,-\,8\lambda/(\Delta+\delta)$ for $\bf H$
parallel and perpendicular to the uniaxial $z$ axis, respectively. \cite{AB} In particular, the spin-orbit coupling
results in deviations of the $g$ factor from the spin-only value $g_s\,=\,2$ and in a small anisotropy of $g$.

Numerically solving the Hamiltonian (1) provides the relative energies of the spin states of the Ni$_4$-cluster as a function of the external
magnetic field. Using the experimentally determined values for the $g$ factor (2.11) and the {ZFC gap $\Delta_1\,=\,6.7$\,K} we performed a
reanalysis of the magnetic susceptibility data {$\chi(T)$} using the spin Hamiltonian described above. This is in particular necessary since the
$D$-value reported in Ref.~\onlinecite{Demeshko05} for this compound was unrealistically large ($D\,=\,-\,50.4$\,K). The present combined analysis
(Fig.~\ref{chi}) of $\chi(T)$ and the ESR data results in magnetic exchange constants $J_1$\,=\,\vl{37.3}\,K (AF coupling), $J_2$\,=\,-\,\vl{57.8}\,K
(FM), a diagonal exchange of $J_3$\,=\,\vl{37.1}\,K (AF) {and yields a realistic value of the single-ion anisotropy parameter $D\,=\,\vl{-}4.8$\,K.}

\begin{figure}
    \includegraphics[width=0.7\columnwidth,angle=-90,clip]{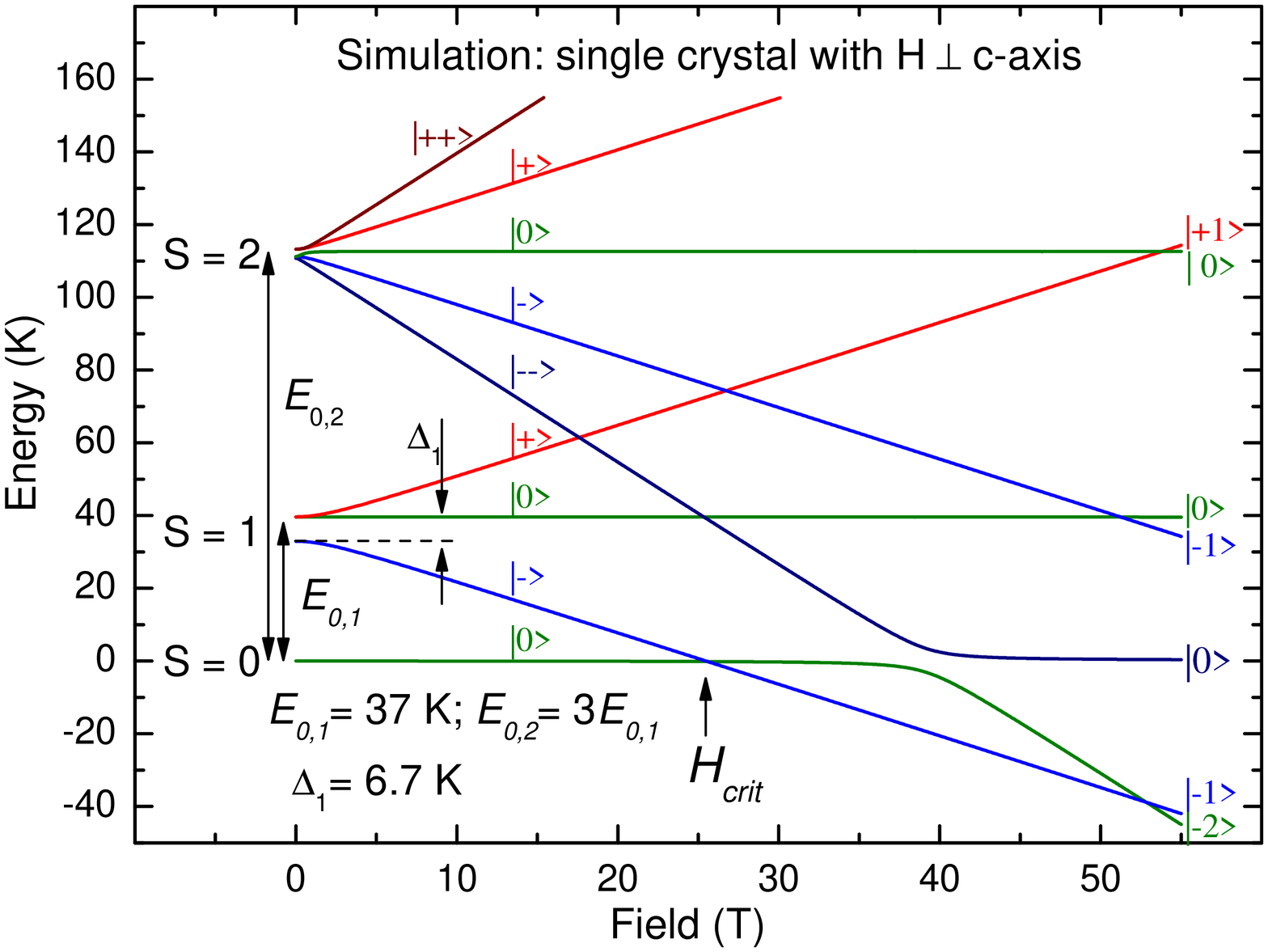}
    \caption{(Color online)Low lying energy levels, calculated for a single crystal
    with the magnetic field perpendicular to the z-axis: the triplet $S\,=\,1$ and the quintuplet $S\,=\,2$ are well separated
    from the $S\,=\,0$ ground state by an activation energy of {$E_{0,1}\,=\,\vl{37}$ K and $E_{0,2}\,\simeq\,3\,E_{0,1}$},
    respectively. Both
    multiplets exhibit a zero field splitting due to the anisotropic CF. Note the level crossing at zero energy
    around $25$\,T which yields the change of the ground state.}
\label{States}
\end{figure}

The coupling of four spins $S_{Ni}\,=\,1$ in the cluster yields an extensive spectrum of $(2S_{Ni}+1)^4 = 81$ individual spin state energies. For
simplicity only the states of lowest energy are shown in Fig.~\ref{States}. The ground state of the quadrumer is rendered a spin singlet {$S\,=\,0$},
which is well separated from the $S$=1 triplet and the $S$=2 quintuplet. The separation energy {$E_{0,i}$, amounts to $E_{0,1}\,\simeq\,\vl{37}$\,K
in the case of the singlet - triplet gap, and $E_{0,2}\,\simeq\,3\,E_{0,1}$ for the singlet - quintuplet gap, respectively.} Remarkably, the ground
state is a spin singlet state although the net exchange is negative { $J\,=\,2\,\sum_{i=1,2,3} J_i = \vl{-33.2}$\,K}, i.e. ferromagnetic. In the
framework of our minimal model this can be understood by considering the hierarchy of the coupling parameters, i.e. by assuming that due to $J_2$ two
FM dimers are formed, for instance pairs of ions (1,3) and (2,4) in Fig.~\ref{ExchangePaths}, which are coupled antiferromagnetically through $J_1$
and $J_3$.

\subsection{Assignment of ESR Modes}

It is straightforward to assign the observed resonance modes in the ESR spectrum to the spin-flip transitions between different energy states of the
Ni$_4$-complex introduced in the previous section. While comparing the experimental results (Fig.~\ref{Branches}) with the numerical results in
Fig.~\ref{States} one should take into account that in a powder spectrum due to the averaging of the different orientations of microcrystallites the
most spectral weight is caused by the particles with the uniaxial CF axis close to the perpendicular direction to the external magnetic field. This
yields the field dependence of the spin state energies shown in Fig.~\ref{States}. The absence of the ESR signals associated with bulk Ni spins at
low temperatures for excitation frequencies $\nu\,\leq\,739$\,GHz and the occurrence of four modes in the spectrum at elevated temperatures suggests
that all resonances in this frequency domain are due to the transitions within thermally populated spin multiplets. The comparison with the numerical
results suggests that resonance branches 1, 2 and 4 depicted in Fig.~\ref{Branches} are due to excitations within the $S\,=\,1$ multiplet
(Fig.~\ref{States}), whereas peak 3 is due to spin resonance in the $S\,=\,2$ quintuplet, lying higher in energy than the triplet. Resonance 1 is a
so-called forbidden transition between levels labelled as $\left|\,-\,\right>$ and $\left|\,+\,\right>$ in Fig.~\ref{States} which in small magnetic
fields are the mixture of $\left|\,-\,1\,\right>$ and $\left|\,+\,1\,\right>$ spin states due to the CF effect and the spin-orbit coupling. \cite{AB}
Therefore the resonance transition between these levels may have an appreciable intensity. Since both $\left|\,-\,\right>$ and $\left|\,+\,\right>$
states change their energy approximately as $-g\mu_BH$ and $+g\mu_BH$, respectively, branch 1 in the $\nu/H^{res}$ diagram in Fig.~\ref{Branches} has
a much stronger slope as compared to other branches.

There is a finite energy gap {$\Delta_1$} in zero magnetic field between the doublet
$\left|\,\vl{+}\,\right>$,
$\left|\,0\,\right>$  and the singlet
$\left|\,\vl{-}\,\right>$
states of the triplet (Fig.~\ref{States}) which is caused by the uniaxial CF anisotropy and the spin-orbit coupling. It can be directly derived from
the intercept of the $\nu(H^{res}_1)$ dependence with the frequency axis. {From the numerical solution of the Hamiltonian~(1) one obtains the
single-ion anisotropy parameter $D\,\simeq\,\vl{-}0.72\,\Delta_1\,=\,\vl{-}4.8$\,K}. Branch 2 having the same frequency offset by the extrapolation
to zero magnetic field has an appreciably weaker field dependence which allows to identify it as a transition between the $\left|\,0\,\right>$ and
$\left|\,\vl{-}\,\right>$
 states within the
$S\,=\,1$ multiplet (Fig.~\ref{States}). From the slope of the $\nu(H^{res}_2)$ dependence one obtains the $g$ factor $g\,=\,2.11\,\pm\,0.01$ which
is within the typical range of Ni $g$ factors. \cite{AB} In contrast to branches 1 and 2, no frequency offset can be found within experimental error
for branch 4 which allows to assign it to a transition between $\left|\,0\,\right>$ and
$\left|\,\vl{+}\,\right>$
 states. The slope of the
$\nu(H^{res}_4)$ dependence yields $g\,=\,2.08\,\pm\,0.01$, similar to the results for branch 2.

In contrast to peaks 1, 2 and 4, which follow a common activation behavior with maxima around $30$ K, resonance peak 3 shows a broad maximum at
roughly $150$ K (Fig.\ref{TempDep}a). This observation implies that peak 3 is due to spin resonance at higher energies, i.e. in the $S=2$ multiplet.
Remarkably, no frequency offset can be revealed in the dependence $\nu(H^{res}_3)$ in agreement with the solution of Hamiltonian (1) which yields a
much smaller zero field splitting of the spin states within the $S\,=\,2$ multiplet as compared with the $S\,=\,1$ multiplet
(Fig.~\ref{Branches}).~\cite{bencini} In the $S\,=\,2$ case the strongest resonance transitions
$\left|\,--\,\right>\,\leftrightarrow\,\left|\,-\,\right>$, $\left|\,-\,\right>\,\leftrightarrow\,\left|\,0\,\right>$,
$\left|\,0\,\right>\,\leftrightarrow\,\left|\,+\,\right>$ and $\left|\,+\,\right>\,\leftrightarrow\,\left|\,++\,\right>$ occur at a fixed excitation
frequency at almost the same value of the resonance field for all transitions. Surprisingly a somewhat larger $g$ factor amounting to
$g\,=\,2.18\,\pm\,0.01$ has been estimated from the $\nu(H^{res}_3)$ slope which is not expected in the framework of Hamiltonian (1). A possible
reason for a discrepancy could be a small experimentally unresolvable zero field splitting of the spin states which however might affect the
lineshape of the total signal from the $S\,=\,2$ multiplet and cause an error in the determination of the resonance field.

\subsection{Ground State in Strong Magnetic Fields }

The calculation of the $H$ dependence of the spin state energies from Hamiltonian (1) with the experimentally determined set of parameters suggests
that the energy of the mixed $\left|\,-\,\right>$ state of the $S\,=\,1$ multiplet drops below the ground state singlet $S\,=0$ at a field of
$\sim\,25$\,T (Fig.~\ref{States}). In fact, in such strong field the quantization axis is defined not by the crystal field anymore but is given by
the direction of the magnetic field. \cite{AB} Therefore the mixed $\left|\,-\,\right>$ state transforms into the $\left|\,-\,1\,\right>$ state
characterized by a single spin quantum number $S_z\,=\,-1$.  Such a level crossing should turn the Ni$_4$-complex to a new magnetic ground state.
Compelling evidence for this peculiar phenomenon comes from the observation of a strong signal within branch 2 at low temperatures in a pulsed field
ESR experiment with the highest excitation frequency of 1017\,GHz (Figs.~\ref{PulseEPR1THzSpec} and \ref{Branches}). As it follows from the
$\nu(H^{res}_2)$ dependence, for $\nu\,=1017$\,GHz the resonance field $H^{res}_2$ exceeds the level crossing point. Therefore, the resonance
excitation occurs from the new magnetic ground state which explains its observation at a low temperature of 4.2\,K and the decrease of its intensity
with increasing $T$. Still a thermally activated resonance within the $S\,=\,2$ multiplet can be observed (branch 3) at elevated temperatures whereas
resonances 1 and 2 are not visible anymore. The former is completely forbidden in strong magnetic fields which turn the mixed $\left|\,-\,\right>$
and $\left|\,+\,\right>$ states into the pure spin states $\left|\,-\,1\,\right>$ and $\left|\,+\,1\,\right>$, respectively. The latter having a much
smaller intensity as compared to resonance 3 cannot be resolved probably due to a limited sensitivity of the equipment at such high frequencies.

The observation of the strong change of the static magnetization around $H_{crit}\,\sim\,25$\,T vigorously supports the level crossing scenario
suggested by the ESR data. The $M(H)$ data give evidence for a strongly magnetic state above 25\,T which is characterized by a large value of the
magnetic susceptibility of the order of $10^{-2}$\, emu/mole Ni. Surprisingly, one finds an appreciable linear increase of $M$ not only above
$H_{crit}$ but also in the low field regime (Fig.~\ref{PulsedFieldMagExp}). Such a linear contribution to $M$ visible in a field range up to 20\,T
cannot be explained by paramagnetic impurities. Furthermore, magnetic coupling between the NiAz molecules in the solid phase is obviously much
smaller than the intramolecular exchange. Thus strong long-range AF correlations between Ni spins in different clusters which could yield the linear
increase of $M(H)$ at $H\,<\,H_{crit}$ are very improbable. In fact, such a linear behavior of the magnetization is reminiscent of the Van Vleck
polarization paramagnetism which is due to the quantum mechanical mixing of different spin states by the magnetic field. However, the experimentally
found values of the linear contribution to $M(H)$ are appreciably larger as compared with the typical Van Vleck susceptibilities which are of the
order of $10^{-4}$\,emu/mole. Therefore, one has to think of a different mechanism which can provide a significant admixture of the high energy
magnetic multiplets to a ground state singlet in a magnetic field. As a possible candidate one may suggest the anisotropic magnetic exchange between
the Ni centers, e.g. the Dzyaloshinsky-Moriya (DM) interaction \cite{DM} which is in principal allowed by symmetry in NiAz. \vl{Indeed}, recent
theoretical calculations reveal an important role of the DM term for the microscopic description of the magnetization of paramagnetic molecular
clusters. \cite{Konstant02} To examine the relevance of the DM interaction  to the case of NiAz theoretical modelling beyond the framework of
Hamiltonian (1) is highly desirable. \vl{In particular, in the presence of an anisotropic interaction one may expect the occurrence of an
anti-crossing energy gap between the $S\,=\,0$ singlet state and the $\left|\,-1\,\right>$ state of the $S\,=\,1$ triplet near the level crossing
point at $H_{crit}$. \cite{Konstant02} Experimentally such a gap yields a hysteresis behavior of the static magnetization owing to nonequilibrium
processes in the spin system which depends on the sweep rate of the magnetic field $\partial H/\partial t$ (see e.g.
Ref.~\onlinecite{Chiorescu00,Waldmann02}). In our pulsed field magnetization experiment the rates $\partial H/\partial t$ are very high and different
for the ascending and descending fields, of the order of 5000 and 500\,T/min, respectively. In this regime we find no hysteresis between the up-
and down sweeps, the observation which calls for a theoretical modelling of the spin-relaxation processes in NiAz near the critical level-crossing
points. The opening of an anti-crossing gap around $H_{crit}$, e.g. owing to the DM interaction, should broaden the spin state transition.
\cite{Konstant02} This may explain a large width $\sim\,10$\,T of the magnetization step of NiAz at $H_{crit}\,\sim\,25$\,T
(Fig.~\ref{PulsedFieldMagExp}). The energy diagram in Fig.~\ref{States} predicts the second level crossing at a field of about 53\,T. Though the
increase of the derivative magnetization above $H\,\sim\,40$\,T which gets even stronger at $H\,>\,50$\,T
(inset of Fig.~\ref{PulsedFieldMagExp}) can be considered as a signature of the second
crossing, no well defined step can be observed in the $M(H)$ curve. Possibly, as suggested by theoretical calculations in
Ref.~\onlinecite{Konstant02}, for particular orientations of the DM vector the width of the magnetization step may increase appreciably with
increasing $H_{crit}$. Thus the experimentally available field range limited to 55\,T might not be sufficient for the observation of the second
strongly smeared out magnetization step.}

\section{Conclusion}

In summary, we have investigated the magnetic properties of a novel quadrumer Ni(II) complex with the central $\mu_4$-azide bridge by ESR and
magnetization measurements in magnetic fields up to 55\,T. Four resonance modes have been found in the ESR spectrum and studied as a function of
temperature and excitation frequency. With experimentally determined parameters of the spin Hamiltonian, such as single ion anisotropy and $g$
factor, the low energy spectrum of the spin states has been calculated. Remarkably, the calculated dependence of the energies on the magnetic field
suggests a change of the ground state from a non magnetic singlet state to a magnetic one in strong magnetic fields due to the level crossing at
$H_{crit}\,\sim\,25$\,T. Indeed, such tuning of the ground state by application of a strong magnetic field has been confirmed in the ESR and
magnetization experiments.

\section{Acknowledgments}

Financial support by the Deutsche Forschungsgemeinschaft through SPP~1137 ''Molecular Magnetism''
 grant KL 1086/6-1 is gratefully acknowledged. Experiments of C.G. and
V.K. in Toulouse were supported by the EuroMagNET consortium of the European Union through contact FP6 R113-CT2004-506239. The work of R.K. in
Toulouse was supported by the DFG through grant KL 1824/1-1.

\end{document}